\documentclass[twoside,english,doublecol]{epl2}
\usepackage[T1]{fontenc}
\usepackage[latin9]{inputenc}
\usepackage{xcolor}
\usepackage{pdfcolmk}
\usepackage{babel}
\usepackage{array}
\usepackage{units}
\usepackage{multirow}
\usepackage{amsmath}
\usepackage{amssymb}
\usepackage{graphicx}
\usepackage{esint}
\PassOptionsToPackage{normalem}{ulem}
\usepackage{ulem}
\usepackage[unicode=true,pdfusetitle,
 bookmarks=true,bookmarksnumbered=false,bookmarksopen=false,
 breaklinks=false,pdfborder={0 0 1},backref=false,colorlinks=false]
 {hyperref}

\makeatletter

\providecommand{\tabularnewline}{\\}
\providecolor{lyxadded}{rgb}{0,0,1}
\providecolor{lyxdeleted}{rgb}{1,0,0}

 \@ifundefined{textcolor}{}
 {%
  \definecolor{BLACK}{gray}{0}
  \definecolor{WHITE}{gray}{1}
  \definecolor{RED}{rgb}{1,0,0}
  \definecolor{GREEN}{rgb}{0,1,0}
  \definecolor{BLUE}{rgb}{0,0,1}
  \definecolor{CYAN}{cmyk}{1,0,0,0}
  \definecolor{MAGENTA}{cmyk}{0,1,0,0}
  \definecolor{YELLOW}{cmyk}{0,0,1,0}
  }

\usepackage{ulem}
\newcommand{\CC}[1]{}
\newcommand{\bS}{\begin{subequations}}
\newcommand{\eS}{\end{subequations}}

\makeatother

\begin{document}
\global\long\def\myvec#1{\mathbf{#1}}
\global\long\def\aseq{\simeq}
\global\long\def\Lp{L_{\parallel}}
\global\long\def\Ls{L_{\perp}}
\global\long\def\Tc{T_{\mathrm{c}}}
\global\long\def\bH{\beta\mathcal{H}}
\global\long\def\para{\parallel}
\global\long\def\sx{\perp_{1}}
\global\long\def\sy{\perp_{2}}
\global\long\def\B{\mathrm{b}}
\global\long\def\db{d}

\global\long\def\Se{\textrm{1d}}
\global\long\def\Sz{2\mathrm{d}}
\global\long\def\Sd{3\mathrm{d}}

\global\long\def\Szb{\textrm{2d}_{\B}}
\global\long\def\Sdb{\textrm{3d}_{\B}}

\global\long\def\See{1{+}1\mathrm{d}}
\global\long\def\Sze{2{+}1\mathrm{d}}
\global\long\def\Sez{1{+}2\mathrm{d}}
\global\long\def\Sed{1{+}3\mathrm{d}}

\pacs{05.70.Ln}{}
\pacs{68.35.Af}{}
\pacs{05.50.+q}{}

\title{Sheared Ising models in three dimensions}

\author{Alfred Hucht and Sebastian Angst}

\institute{Fakultät für Physik, Universität Duisburg-Essen, D-47048 Duisburg}

\date{\today}

\abstract{The nonequilibrium phase transition in sheared three-dimensional
Ising models is investigated using Monte Carlo simulations in two
different geometries corresponding to different shear normals. We
demonstrate that in the high shear limit both systems undergo a strongly
anisotropic phase transition at exactly known critical temperatures
$\Tc$ which depend on the direction of the shear normal. Using dimensional
analysis, we determine the anisotropy exponent $\theta=2$ as well
as the correlation length exponents $\nu_{\para}=1$ and $\nu_{\perp}=1/2$.
These results are verified by simulations, though considerable corrections
to scaling are found. The correlation functions perpendicular to the
shear direction can be calculated exactly and show Ornstein-Zernike
behavior.}

\maketitle

\section{Introduction}

While the occurrence of nonequilibrium phase transitions is ubiquitous
in nature, its investigation in the framework of nonequilibrium statistical
mechanics is intricate and restricted to a few simple models, like
the driven lattice gas (DLG) \cite{KatzLebowitzSpohn83,SchmittmannZia95,Zia10}
or, recently, to the driven two-dimensional Ising model \cite{KadauHuchtWolf08}.
In this model the system is cut into two halves parallel to one axis
and moved along this cut with the velocity $v$. The model exhibits
energy dissipation and subsequently friction due to spin correlations,
which also occurs in a suitable Heisenberg model \cite{MagieraBrendelWolfNowak09,MagieraWolfBrendelNowak09,MagieraBrendelWolfNowak11,MagieraAngstHuchtWolf11}
and, of interest for the current context, undergoes a nonequilibrium
(surface) phase transition. The latter has been investigated analytically
and with Monte Carlo (MC) simulations for various geometries \cite{Hucht09}.
Since then, this model has been generalized to the driven Potts models
\cite{IgloiPleimlingTurban11}, and finite-size effects were calculated
analytically in the driven Ising chain \cite{Hilhorst11}.

A lot of similarities and comparable critical behavior between the
Ising model with friction and the very famous and well investigated
DLG have been found \cite{AngstHuchtWolf12}. Both models are characterized
by a critical temperature $\Tc$, which increases with the \textcolor{black}{driving
strength}, the field and the shift or shear velocity $v$, respectively,
and saturates in the high driving limit. For diverse geometries of
the Ising model with friction, the critical temperature has been calculated
analytically for $v\rightarrow\infty$ \cite{Hucht09}. 

Moreover it was discovered that the DLG and two-dimensional sheared
Ising systems with non-conserved order parameter \cite{SaraccoGonnella09,WinterVirnauHorbachBinder10,AngstHuchtWolf12}
show strongly anisotropic critical behavior, with direction dependent
correlation length exponents $\nu_{\para}$ and $\nu_{\perp}$. For
the $\Sz$ and $\See$ geometry of the Ising model with shear the
same exponents $\nu_{\parallel}=3/2$ and $\nu_{\perp}=1/2$ \cite{AngstHuchtWolf12}
as in the two-dimensional DLG have been determined. Additionally finite
velocities $v$ have been studied and it was found that for all finite
$v$ the $\Sz$ and $\See$ model cross-over from isotropic Ising
like behavior to strongly anisotropic mean-field behavior in the thermodynamic
limit, demonstrating that the external drive is a relevant perturbation.

In the following we extend the investigations to three-dimensional
models in two different shear geometries and focus on the high shear
velocity limit $v\rightarrow\infty$. Both represent three-dimensional
sheared models and they are therefore experimentally accessible in
the framework of sheared binary liquids \cite{CrossHohenberg93,Hashimoto95,Onuki97,Migler01},
albeit the order parameter is not conserved here. Using dimensional
analysis, we predict the correlation length exponents for arbitrary
dimension $\db$. These predictions are verified by simulations, however
we find strong corrections to scaling at small system sizes.

\section{Model}

\begin{figure}
\begin{centering}
\includegraphics[scale=0.35]{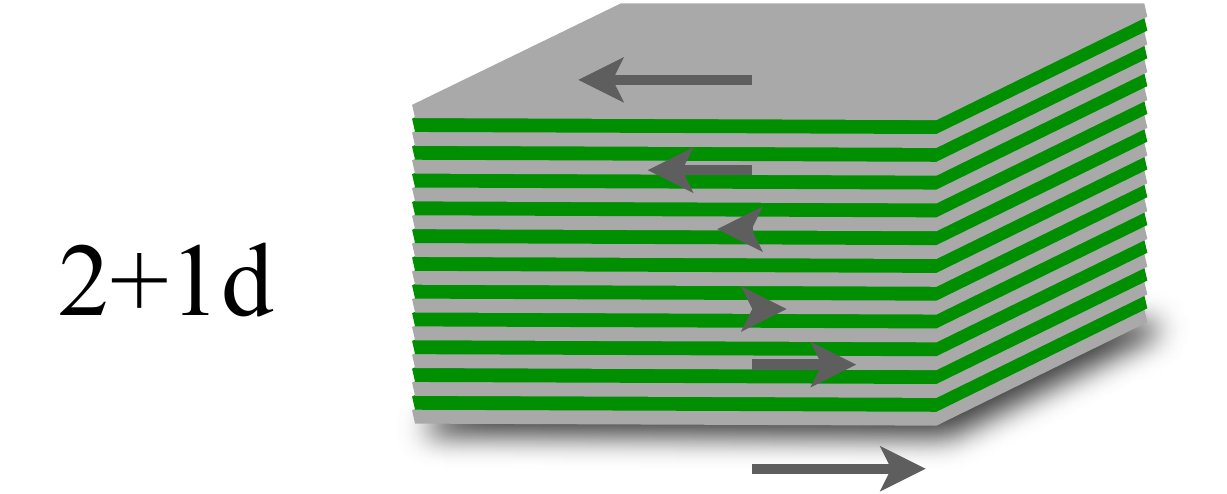}\includegraphics[scale=0.35]{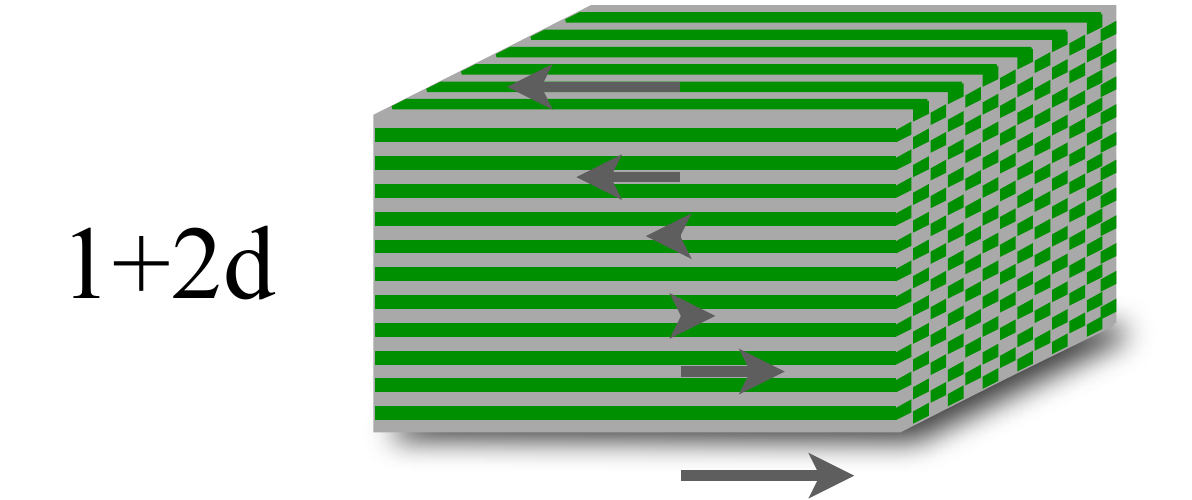}
\par\end{centering}

\caption{Sketches of the systems considered in this work. On the left hand
side the 2+1d system and on the right hand side the 1+2d system is
shown. The gray regions represent the magnetic systems and the green
(dark) regions are the moving boundaries, while the arrows indicate
the motion of the subsystems. \label{fig:model}}
\end{figure}

The systems considered in this work are denoted $\Sze$ and $\Sez$
and are shown in Fig.\,\ref{fig:model}, for a classification see
Ref.~\cite{Hucht09}. In the $\Sze$ geometry shear is applied such
that two-dimensional Ising models are moved relative to their upper
(lower) neighboring layer with velocity $v$ ($-v$) along an axis.
In the following we denote the direction parallel to the shear with
$\para$, the direction perpendicular to the planes with $\sx$ and
the inplane direction perpendicular to the shear direction with $\sy$.
The model contains $L_{\sx}\times L_{\sy}\times\Lp$ spins (lattice
sites), where we choose $L_{\sx}=L_{\sy}=:L_{\perp}$ throughout this
work, and periodic boundary conditions are applied in all directions.
The shear velocity $v$ corresponds to a shear rate, which is often
denoted as $\dot{\gamma}$ \cite{SaraccoGonnella09,WinterVirnauHorbachBinder10}.
Using the notation $(\sx\,\sy\,\para)$ for directions, the shear
is in $(001)$-direction and the shear normal is in $(100)$-direction. 

A finite shear velocity $v$ is implemented by shifting neighboring
layers $v$ times by one lattice constant during one MC step (for
details see \cite{KadauHuchtWolf08,Hucht09}). A simplification of
the implementation is yielded by reordering the couplings between
moved layers instead, and by introducing a time-dependent displacement
$\Delta(t)=vt$ we get the Hamiltonian
\begin{eqnarray}
\bH(t) & = & -\sum_{k=1}^{L_{\sx}\vphantom{\Lp}}\sum_{l=1}^{L_{\sy}\vphantom{\Lp}}\sum_{m=1}^{\Lp}\sigma_{klm}\Big(K_{\para}\sigma_{k,l,m+1}+\nonumber \\
 &  & {}+K_{\sy}\sigma_{k,l+1,m}+K_{\sx}\sigma_{k+1,l,m+\Delta(t)}\Big),\label{eq:Hamiltonian_2+1d}
\end{eqnarray}
where $K_{\mu}=\beta J_{\mu}$ is the reduced nearest neighbor coupling
with $\mu=\{\sx,\sy,\para\}$, and $\beta=1/k_{\mathrm{B}}T$. In
the following we concentrate on the infinite shear velocity limit
$v\to\infty$, which can easily be implemented by choosing $1\le\Delta(t)\leq L_{\parallel}$
randomly. In this limit an analytical calculation \cite{Hucht09}
yield the equation
\begin{equation}
\chi_{\mathrm{eq}}^{(0)}(K_{\mathrm{c},\parallel})f\tanh K_{\mathrm{c},\perp}=1\label{eq:chi_Tc}
\end{equation}
from which we can determine the critical temperature, where $\chi_{\mathrm{eq}}^{(0)}$
is the zero field equilibrium susceptibility of the subsystems moved
relative to each other and $f$ the number of fluctuating adjacent
fields. Here $\chi_{\mathrm{eq}}^{(0)}$ of the two-dimensional Ising
model is required, which has been calculated to higher than $2000^{\mathrm{{th}}}$
order by an polynomial algorithm \cite{Boukraa08}. Using $f=2$ and
$J_{\para}=J_{\sx}=J_{\sy}=1$ we get 
\begin{equation}
\Tc^{\Sze}(\infty)=5.2647504145147435505980\ldots\,.\label{eq:Tc_2+1d}
\end{equation}

The second considered geometry $\Sez$ is similar to the previous
case, but now the shear normal is in the $(110)$-direction. As a
consequence, all four perpendicular coupling partners of a spin $\sigma$
are in neighboring shear planes. The corresponding Hamiltonian reads
\begin{eqnarray}
\bH(t) & = & -\sum_{k=1}^{L_{\sx}\vphantom{\Lp}}\sum_{l=1}^{L_{\sy}\vphantom{\Lp}}\sum_{m=1}^{\Lp}\sigma_{klm}\Big(K_{\para}\sigma_{k,l,m+1}+\nonumber \\
 &  & {}+K_{\perp}[\sigma_{k,l+1,m+\Delta(t)}+\sigma_{k+1,l,m+\Delta(t)}]\Big),\label{eq:Hamiltonian_1+2d}
\end{eqnarray}
where $K_{\sx}=K_{\sy}=:K_{\perp}$. For $v\to\infty$ we set $f=4$
and use $\chi_{\mathrm{eq}}^{(0)}(K_{\mathrm{c},\parallel})=e^{2K_{\mathrm{c},\parallel}}$
from the one-dimensional Ising model in Eq.~\eqref{eq:chi_Tc} to
get, for $J_{\para}=J_{\perp}=1$, the critical temperature 
\begin{equation}
\Tc^{\Sez}(\infty)=\frac{2}{\ln[\frac{1}{8}(5+\sqrt{41})]}=5.642611138\ldots,\label{eq:Tc_1+2d}
\end{equation}
which notably is different from Eq.~\eqref{eq:chi_Tc}. Hence the
critical temperature depends on the direction of the shear normal. 

In MC simulations of nonequilibrium models the critical temperature
often depends on the used acceptance rates \cite{KwakLandauSchmittmann04}.
It has been shown that the multiplicative rate \cite{Hucht09} 
\begin{equation}
p_{\mathrm{\mathrm{{flip}}}}(\Delta E)=e^{-\frac{\beta}{2}(\Delta E-E_{\mathrm{{min}}})}
\end{equation}
with the energy change $\Delta E$ and the minimal energy change $\Delta E_{\mathrm{{min}}}=\mathrm{{min}\left\{ \Delta E\right\} }$
must be used in order to reproduce the critical temperatures Eqs.~(\ref{eq:Tc_2+1d},
\ref{eq:Tc_1+2d}).

\section{Anisotropic scaling}

Our aim is to proof that both models exhibit a strongly anisotropic
phase transition and calculate the corresponding exponents. Such a
phase transition is characterized by bulk correlation lengths $\xi_{\mu}$
diverging with direction dependent critical exponents $\nu_{\mu}$
at criticality %
\footnote{Throughout this work the symbol $\aseq$ means ``asymptotically equal''
in the respective limit, e.g., $f(L)\aseq g(L)\Leftrightarrow\lim_{L\rightarrow\infty}f(L)/g(L)=1$
.%
}, 
\begin{equation}
\xi_{\mu}(t)\stackrel{{\scriptscriptstyle t>0}}{\aseq}\hat{\xi}_{\mu}t^{-\nu_{\mu}},\label{eq:def_xi}
\end{equation}
with direction $\mu=\{\sx,\sy,\para\}$, amplitude $\hat{\xi}_{\mu}$,
and reduced critical temperature $t=T/\Tc-1$. Usually one defines
the anisotropy exponent $\theta=\nu_{\para}/\nu_{\perp}$, which is
$\theta=1$ for isotropic scaling and $\theta\ne1$ for strongly anisotropic
scaling \cite{Selke88,BinderWang89,SchmittmannZia95,Hucht02a,AlbanoBinder12}.
As mentioned above, the phase transitions of the Ising model with
friction in the $\Sz$ and the $\See$ geometry become strongly anisotropic
for $v>0$ in the thermodynamic limit, with $\theta=3$ \cite{AngstHuchtWolf12}.

In Ref.~\cite{AngstHuchtWolf12} it was shown that the application
of a stripe geometry $L_{\perp}\rightarrow\infty$ with finite $\Lp$
is an appropriate way to determine the anisotropy exponent and subsequently
the correlation length exponents. Hence we measure the perpendicular
correlation function
\begin{equation}
G_{\perp}(\Lp;\myvec r_{\perp})=\langle\sigma_{000}\sigma_{r_{\sx},r_{\sy},0}\rangle
\end{equation}
at the critical point $\Tc$, from which we can determine the correlation
lengths $\xi_{\mu}$ with $\mu=\left\{ \sx,\sy\right\} $ as shown
below (in the following the index $\mu$ only represents the perpendicular
directions $\sx$ and $\sy$). Note that by symmetry $G_{\perp}(\Lp,r_{\sx})=G_{\perp}(\Lp,r_{\sy})$
for the $\Sez$ system. From $\xi_{\mu}$ we can then determine $\theta$
using the relation \cite{HenkelSchollwoeck01,Hucht02a} 
\begin{equation}
\xi_{\mu}(\Lp)\aseq A_{\mu}L_{\parallel}^{1/\theta}.\label{eq:xi(Lp)}
\end{equation}
 The above-mentioned stripe geometry is a film geometry in three dimensions,
and we choose $\Ls/\xi_{\perp}(\Lp)\gtrsim10$ sufficient for our
purpose \cite{AngstHuchtWolf12}.

\section{Dimensional analysis}

For $v\to\infty$ it was shown in Ref.~\cite{Hucht09} that the 1+1d
model can be mapped onto an \emph{equilibrium} system consisting of
one-dimensional chains that only couple via fluctuating magnetic fields.
Due to the stripe geometry with short length $\Lp$ and the periodic
boundary conditions in parallel direction the magnetization $m(\myvec x)$
with $\myvec x=(\myvec x_{\perp},x_{\para})$ is homogeneous in this
direction, and parallel correlations are irrelevant. Hence we can
use the zero mode approximation in this direction, leading to an order
parameter $m=m(\myvec x_{\perp})$ only. 

The resulting Ginzburg-Landau-Wilson (GLW) Hamiltonian 
\begin{equation}
\beta\mathcal{H}=\Lp\int\mathrm{d}x_{\perp}^{\db-1}\left(\frac{t}{2}m^{2}+\frac{1}{2}(\nabla m)^{2}+\frac{u}{4!}m^{4}\right)\label{eq:Hamiltonian_GLW-d}
\end{equation}
can, however, not be mapped onto a Schrödinger equation for systems
with $\db>2$ as done in Ref.~\cite{AngstHuchtWolf12}, as the $(\db{-}1)$-dimensional
integral cannot be interpreted as a time integral. Instead we use
dimensional analysis in order to predict the critical exponents: starting
from the GLW Hamiltonian \eqref{eq:Hamiltonian_GLW-d} in $\db$ dimensions
we eliminate $\Lp$ with the substitution \bS\label{eq:Substitution}
\begin{eqnarray}
m & \rightarrow & \tilde{m}\,\Lp^{-1/(5-\db)}\label{eq:rescal_m}\\
\myvec x_{\perp} & \rightarrow & \tilde{\myvec x}\,\Lp^{1/(5-\db)}\label{eq:rescal_x}\\
t & \rightarrow & \tilde{t}\,\Lp^{-2/(5-\db)}\label{eq:rescal_t}
\end{eqnarray}
\eS to get the $(\db{-}1)$-dimensional Hamiltonian 
\begin{equation}
\beta\mathcal{H}=\int\mathrm{d}\tilde{x}^{\db-1}\left(\frac{\tilde{t}}{2}\tilde{m}^{2}+\frac{1}{2}(\nabla\tilde{m})^{2}+\frac{u}{4!}\tilde{m}^{4}\right),\label{eq:Hamiltonian_GLW-s}
\end{equation}
with $\tilde{m}=\tilde{m}(\tilde{\myvec x})$. From Eqs.~(\ref{eq:Substitution}b,c)
we directly read off the exponents 
\begin{equation}
\theta=5-\db,\qquad\nu_{\para}=\frac{5-\db}{2},\quad\Rightarrow\quad\nu_{\perp}=\frac{1}{2},\label{eq:exponents_d}
\end{equation}
reproducing the results for $\db=1$ \cite{Hucht09} and $\db=2$
\cite{AngstHuchtWolf12} and fulfilling the generalized hyperscaling
relation \cite{Binder90} 
\begin{equation}
\nu_{\para}+(\db-1)\nu_{\perp}=2-\alpha
\end{equation}
with $\alpha=0$ \cite{Hucht09,AngstHuchtWolf12}. For our case $\db=3$
we find 
\begin{equation}
\theta=2,\qquad\nu_{\para}=1,\qquad\nu_{\perp}=\frac{1}{2},\label{eq:exponents_3}
\end{equation}
while for $\db\geq4$ we predict isotropic or weakly anisotropic behavior
with $\theta=1$ and $\nu_{\para}=\nu_{\perp}=1/2$, as then the upper
critical dimension $d_{\mathrm{c}}=4$ is reached and the shear becomes
an irrelevant perturbation.

\section{Correlation functions}

The perpendicular correlation function can be calculated from Eq.~\eqref{eq:Hamiltonian_GLW-s}
using a Gaussian approximation, which is valid, since we investigate
the system at the critical temperature of the bulk, which is higher
than the the critical temperature of the studied film geometry. Setting
$u=0$ in Eq.~\eqref{eq:Hamiltonian_GLW-s} and using $\tilde{\xi}\propto\tilde{t}^{-1/2}$
we get the Ornstein-Zernike structure factor
\begin{equation}
\tilde{S}(\tilde{\myvec k})\propto\frac{1}{\tilde{k}{}^{2}+\tilde{\xi}^{-2}}.\label{eq:Sst}
\end{equation}
In our case the dimension is $d=3$, and a Fourier transformation
yields the correlation function 
\begin{equation}
\tilde{G}(\tilde{\myvec r})\propto K_{0}(\tilde{r}/\tilde{\xi}).\label{eq:Gst}
\end{equation}
Using $\tilde{G}\propto\tilde{m}^{2}\propto\Lp^{-1/\nu_{\para}}$
and back-substituting with Eqs.~\eqref{eq:Substitution} gives the
result
\begin{equation}
G(\Lp;\myvec r_{\perp})\propto\Lp^{-1/\nu_{\para}}K_{0}[r_{\perp}/\xi_{\perp}(\Lp)]\label{eq:GsGLW}
\end{equation}
for the perpendicular correlation function of the GLW Hamiltonian
\eqref{eq:Hamiltonian_GLW-d}, with modified Bessel function of the
second kind $K_{0}$.

The 2+1d geometry is weakly anisotropic in perpendicular direction
at least for different couplings $J_{\sx}\neq J_{\sy}$, i.e., the
correlation lengths $\xi_{\sx}$ and $\xi_{\sy}$ have same exponent
$\nu_{\perp}$ but different amplitudes $\hat{\xi}_{\mu}$ \cite{Hucht02a}.
This anisotropy can be removed by the rescaling 
\begin{equation}
l_{\mu}\rightarrow\bar{l}_{\mu}=\frac{l_{\mu}}{A_{\mu}},\label{eq:rescale}
\end{equation}
with amplitude $A_{\mu}$ from Eq.~\eqref{eq:xi(Lp)}. Now the perpendicular
directions are isotropic and we can use Eq.~\eqref{eq:GsGLW} to
get the final result
\begin{equation}
G_{\perp}(\Lp;r_{\mu})\aseq\hat{G}\Lp^{-1/\nu_{\para}}K_{0}[r_{\mu}/\xi_{\mu}(\Lp)]\label{eq:Gs_2d}
\end{equation}
for the two directions $\mu={\sx}$ and $\sy$. Here we already have
back-substituted with Eq.~\eqref{eq:rescale}. Note that especially
in the 2+1d case the amplitude $\hat{G}$ should not depend on the
direction $\mu$.

\section{Results}

\begin{table}[!b]
\begin{centering}
\begin{tabular}{c|c|r@{\extracolsep{0pt}.}l|r@{\extracolsep{0pt}.}l|r@{\extracolsep{0pt}.}l}
Model & $\mu$ & \multicolumn{2}{c|}{$A_{\mu}$} & \multicolumn{2}{c|}{$\hat{G}$} & \multicolumn{2}{c}{$c_{0}$}\tabularnewline
\hline 
\hline 
$\Sez$ & $\perp$ & 0&254(5) & 0&93(1) & 14&(1)\tabularnewline
\multirow{2}{*}{$\Sze$} & $\sx$ & 0&320(5) & 0&85(1) & 12&(1)\tabularnewline
 & $\sy$ & 0&331(5) & 0&85(1) & 12&(1)\tabularnewline
\end{tabular}
\par\end{centering}

\caption{Amplitudes and corrections to scaling parameter $c_{0}$ for both
models.\label{tab:Fit-parameters}}
\end{table}
\begin{figure}
\begin{centering}
\includegraphics[width=1\columnwidth]{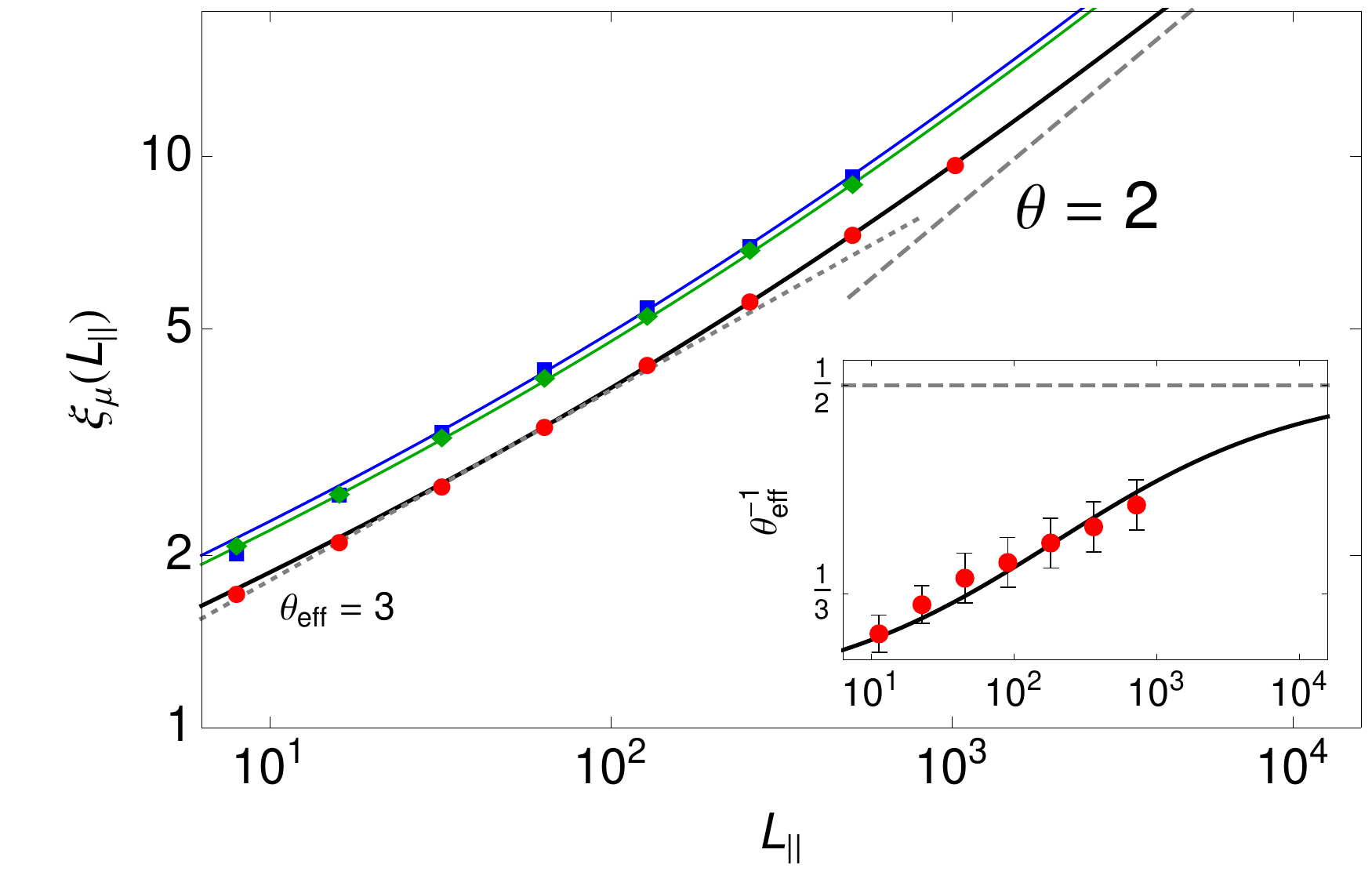}
\par\end{centering}

\caption{Perpendicular correlation lengths $\xi_{\mu}(\Lp)$ for the $\Sez$
geometry (red circles), the $\Sze$ geometry in the $\sx$-direction
(green diamonds) and in the $\sy$-direction (blue squares) at criticality.
The statistical error is smaller than the symbol size. Due to corrections
to scaling small systems have effective anisotropy exponent $\theta_{\mathrm{eff}}\approx3$
(dotted line), which is obtained from the logarithmic derivative and
shown exemplary for system $\Sez$ in the inset. \label{fig:xis_von_Lp}}
\end{figure}

We measured $G_{\perp}(\Lp;r_{\mu})$ at criticality for both models
using extensive Monte Carlo simulations and fitted the results against
Eq.~\eqref{eq:Gs_2d} to get $\xi_{\mu}(\Lp)$ shown in Fig.~\ref{fig:xis_von_Lp}.
As in the 1+1d case we find corrections to scaling for $\Lp\lessapprox300$
which are problematic in these three-dimensional cases as we cannot
simulate systems larger than $\Lp=1024$. Hence we have to introduce
a lattice correction term in the perpendicular correlation length
and improve relation \eqref{eq:xi(Lp)} using the \emph{ansatz }
\begin{equation}
\xi_{\mu}(\Lp)=A_{\mu}(\Lp+c_{0}\Lp^{1/2}+\ldots)^{1/\theta}\label{eq:xi(Lp)_ansatz}
\end{equation}
with $\theta=2$, which gives the best fit to the data. From the numerical
data we find the amplitudes $A_{\mu}$ and $\hat{G}$ as well as the
correction parameter $c_{0}$ listed in Tab.~\ref{tab:Fit-parameters},
and the resulting fit is shown as solid line in Fig.~\ref{fig:xis_von_Lp}.
For large systems the curve approaches the theoretical limit Eq.\,(\ref{eq:xi(Lp)})
with slope $\theta^{-1}=1/2$. Note that for small $\Lp\lesssim64$
we could also find a reasonable data collapse with exponent $\theta_{\mathrm{eff}}=3$
(dotted line).

\begin{figure}
\begin{centering}
\includegraphics[width=1\columnwidth]{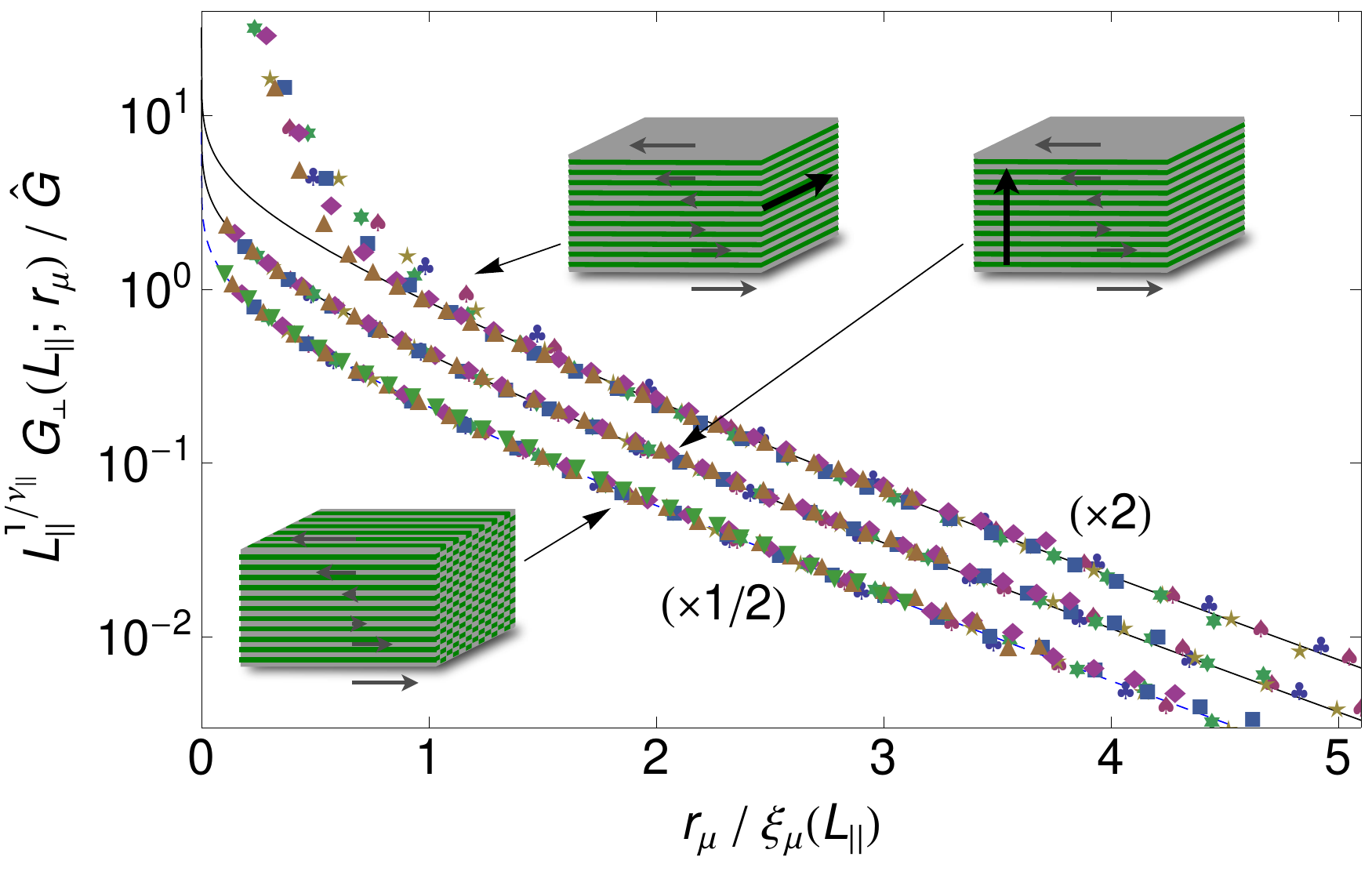}
\par\end{centering}

\caption{Rescaled correlation function $G_{\perp}(\Lp;r_{\mu})$ with $\mu=\{\perp,\sx,\sy\}$
for both models at criticality. We show varying system extensions
$\Lp=\text{\{8,16,32,64,128,256,512,1024\}}$ for both cases. A rescaling
of the $x$-axis with $\xi_{\mu}(\Lp)$ and of the $y$-axis with
$\Lp$ results in an excellent data collapse, verifying $\theta=2$
and $\nu_{\para}=1$. The solid lines represent the calculated Ornstein-Zernike
correlation function, Eq.~\eqref{eq:Gs_2d}. Note that we multiplied
the collapsed data with different factors as indicated in order to
show them in one plot.\label{fig:result}}
\end{figure}

The resulting rescaled correlation functions for both models are presented
in Fig.\,\ref{fig:result}. In all cases the $y$-axis can be rescaled
with $\Lp$ as predicted, without notable corrections. We find a convincing
data collapse onto the mean-field correlation function $K_{0}(r/\xi)$
from Eq.~\eqref{eq:Gs_2d}. For small distances $r_{\sy}=\mathcal{O}(1)$
the correlation function $G_{\perp}(\Lp;r_{\sy})$ differs from Eq.~(\ref{eq:Gs_2d})
due to the inplane nearest neighbor interactions.

Now we comment on the four-dimensional geometry 1+3d, with decouples
to a three-dimensional array of interacting chains, with $f=6$ in
Eq.~\eqref{eq:chi_Tc}. We performed test simulations for system
sizes up to $32^{3}\times32$ and found very strong, possibly logarithmic
corrections to scaling. From the scaling behavior of the available
data we estimate that system sizes $\Lp,\Ls\gtrsim1000$ would be
required to find the correct scaling behavior.

Finally, we extend the dimensional analysis to the general case of
a $d$-dimensional hyper-cubic sheared lattice with $d_{\para}$ driven
dimensions and $d_{\perp}$ perpendicular dimensions. We again must
distinguish between the $d_{\sx}$ dimensions normal to the shear
and $d_{\sy}$ ``inplane'' dimensions without shear motion, with
$d_{\perp}=d_{\sx}+d_{\sy}$. The critical temperature $\Tc$ at infinite
shear velocity $v$ is given by Eq.~\eqref{eq:chi_Tc}, with the
equilibrium zero field susceptibility $\chi_{\mathrm{eq}}^{(0)}$
of the $d_{\mathrm{eq}}$-dimensional system having $f$ fluctuating
fields at each lattice point, where $d_{\mathrm{eq}}=d_{\para}+d_{\sy}$,
and $f=2d_{\sx}$. From a simple generalization of Eq.~\eqref{eq:exponents_d}
we find the exponents
\begin{equation}
\theta=\frac{4-d_{\perp}}{d_{\para}},\qquad\nu_{\para}=\frac{4-d_{\perp}}{2d_{\para}},\qquad\nu_{\perp}=\frac{1}{2},\label{eq:general_exponents}
\end{equation}
fulfilling the hyperscaling relation $d_{\para}\nu_{\para}+d_{\perp}\nu_{\perp}=2$. 

\begin{table*}

\begin{centering}
\begin{tabular}{cc|ccccc|cc|cc|r@{\extracolsep{0pt}.}l}
\multicolumn{2}{c|}{model} & $\, d\,$ & $d_{\para}$ & $d_{\perp}$ & $d_{\sx}$ & $d_{\sy}$ & $\theta$ & $\nu_{\para}$ & $f$ & $d_{\mathrm{eq}}$ & \multicolumn{2}{c}{$\Tc(\infty)/J$}\tabularnewline
\hline 
\multirow{5}{*}{\rotatebox{90}{moved}} & $\Se$ & 1 & 1 & -- & -- & -- & -- & 2 & 1 & 1 & ~2&2691853$\ldots$\tabularnewline
 & $\Sz$ & 2 & 1 & 1 & 0 & 1 & 3 & $\nicefrac{3}{2}$ & 1 & 2 & 4&0587824$\ldots$\tabularnewline
 & $\Sd$ & 3 & 1 & 2 & 0 & 2 & 2 & 1 & 1 & 3 & 5&983835(1)\tabularnewline
 & $\Szb$ & 1 & 1 & -- & -- & -- & -- & 2 & 1 & 2 & 2&6614725$\ldots$ \tabularnewline
 & $\Sdb$ & 2 & 1 & 1 & 0 & 1 & 3 & $\nicefrac{3}{2}$ & 1 & 3 & 4&8(1)\tabularnewline
\hline 
\multirow{4}{*}{\rotatebox{90}{sheared}} & $\See$ & 2 & 1 & 1 & 1 & 0 & 3 & $\nicefrac{3}{2}$ & 2 & 1 & 3&4659074$\ldots$\tabularnewline
 & $\Sze$ & 3 & 1 & 2 & 1 & 1 & 2 & 1 & 2 & 2 & 5&2647504$\ldots$\tabularnewline
 & $\Sez$ & 3 & 1 & 2 & 2 & 0 & 2 & 1 & 4 & 1 & 5&6426111$\ldots$\tabularnewline
 & $\Sed$ & 4 & 1 & 3 & 3 & 0 & 1 & $\nicefrac{1}{2}$ & 6 & 1 & 7&728921$\ldots$\tabularnewline
\hline 
\multirow{2}{*}{\rotatebox{90}{mix}} & $\mathrm{\Sz}_{\mathrm{m}}$ & 2 & 2 & -- & -- & -- & -- & 1 & 1 & 1 & 4&0587824$\ldots$\tabularnewline
 & $\Sze_{\mathrm{m}}$ & 3 & 2 & 1 & 1 & 0 & $\nicefrac{3}{2}$ & $\nicefrac{3}{4}$ & 2 & 2 & 5&2647504$\ldots$\tabularnewline
\end{tabular}
\par\end{centering}

\caption{Relevant dimensions, exponents and parameters of the considered models,
as defined in the text. For a classification see \cite{Hucht09}.
\label{tab:dimensions}}
\end{table*}
We conclude with a tabular summary of the found exponents and critical
temperatures $\Tc$ at infinite driving velocity $v$ given in Table~\ref{tab:dimensions},
including two cases denoted ``mix'' where we assumed a suitable
two-dimensional motion of the interacting planes. These systems have
$d_{\para}=2$, but notwithstanding the same $\Tc$ as the corresponding
systems with unidirectional motion at infinite $v$. For the layered
case $2{+}1\mathrm{d}_{\mathrm{m}}$ we predict the exponents $\theta=3/2$
and $\nu_{\para}=3/4$. A test of these predictions is left for future
work.

\section{Conclusion}

We investigated the phase transition of three-dimensional Ising models
with shear and two different shear normals by means of Monte Carlo
simulations. In the limit of infinitely high shear velocity $v$ we
found a critical temperature $\Tc(\infty)$ that depends on the direction
of the shear normal. At criticality, strongly anisotropic diverging
correlation lengths, with exponents $\nu_{\para}=1$ and $\nu_{\perp}=1/2$
occur, leading to an anisotropy exponent $\theta=2$, which confirms
the results of a dimensional analysis of the corresponding Ginzburg-Landau-Wilson
Hamiltonian. Furthermore, the dimensional analysis captures the anisotropy
exponents as well as the correlation length exponents of the previously
studied two-dimensional cases \cite{AngstHuchtWolf12} and the parallel
correlation length exponent of the one-dimensional cases \cite{Hucht09}.
Predictions for two-dimensional shear directions also result from
the dimensional analysis, leading to the exponents $\theta=3/2$ and
$\nu_{\para}=3/4$ in a three-dimensional model. Fluctuations perpendicular
to the shear were shown to be Gaussian, resulting in a correlation
function with Ornstein-Zernike behavior. Additionally, in the case
of the $\Sze$ geometry we found weakly anisotropic perpendicular
correlations. As for $v=0$ the 2+1d and the 1+2d geometry reduce
to the three-dimensional equilibrium Ising model, we expect a cross-over
from this case to strongly anisotropic mean-field behavior similar
to the $\See$ geometry. In Ref.~\cite{AngstHuchtWolf12} an expensive
analysis for finite velocities has been done leading to a crossover
scaling, pointing out that all $v\ne0$ provoke strongly anisotropic
mean-field behavior, which is expected to occur in the current systems
as well. However, we did not proof this in detail, due to the additional
complexity in three-dimensional systems. 
\begin{acknowledgments}
We thank Felix M. Schmidt and D. E. Wolf for very valuable discussions.
This work was supported by CAPES--DAAD through the PROBRAL program
as well as by the German Research Society (DFG) through SFB 616 ``Energy
Dissipation at Surfaces''. 
\end{acknowledgments}
\bibliographystyle{/users/fred/TeX/bst/eplbib}
\bibliography{Physik}

\end{document}